\documentclass{article}
\usepackage{spconf,amsmath,graphicx}
\usepackage{booktabs}
\usepackage{multirow}
\usepackage{xcolor}
\usepackage{url}
\usepackage{siunitx}
\usepackage{enumitem}
\setlist{nosep, leftmargin=14pt}

\title{Loss Design and Architecture Selection for Long-Tailed Multi-Label Chest X-Ray Classification}

\name{Nikhileswara Rao Sulake}
\address{Department of Computer Science and Engineering, \\Rajiv Gandhi University of Knowledge Technologies, Nuzvid, India\thanks{Corresponding Author: nikhil01446@gmail.com}}

\begin{document}
\maketitle

\begin{abstract}
Long-tailed class distributions pose a significant challenge for multi-label chest X-ray (CXR) classification, where rare but clinically important findings are severely underrepresented. In this work, we present a systematic empirical evaluation of loss functions, CNN backbone architectures and post-training strategies on the CXR-LT 2026 benchmark, comprising approximately 143K images with 30 disease labels from PadChest. Our experiments demonstrate that LDAM with deferred re-weighting (LDAM-DRW) consistently outperforms standard BCE and asymmetric losses for rare class recognition. Amongst the architectures evaluated, ConvNeXt-Large achieves the best single-model performance with 0.5220 mAP and 0.3765 F1 on our development set, whilst classifier re-training and test-time augmentation further improve ranking metrics. On the official test leaderboard, our submission achieved 0.3950 mAP, ranking 5th amongst all 68 participating teams with total of 1528 submissions. We provide a candid analysis of the development-to-test performance gap and discuss practical insights for handling class imbalance in clinical imaging settings. Code is available at \url{https://github.com/Nikhil-Rao20/Long_Tail}.

\end{abstract}


\section{Introduction}
\label{sec:intro}

Chest X-rays (CXRs) remain the most widely performed diagnostic imaging examination across the world, serving as the primary tool for detecting a broad range of thoracic abnormalities. However, the distribution of clinical findings in CXR datasets inherently follows a long-tailed pattern, conditions like cardiomegaly and pleural effusion are observed frequently, whereas findings such as pneumothorax and emphysema occur relatively rarely. This severe class imbalance poses a fundamental challenge for deep learning based classification systems, since conventional loss functions tend to get biased towards the head classes, resulting in models that fail to adequately recognise rare but clinically significant pathologies. Further complicating this, CXR interpretation is inherently a multi-label problem as a single patient can simultaneously present multiple co-occurring findings. Unlike single-label tasks common in computer vision, models here must learn complex label co-occurrence patterns whilst handling severe imbalance across all findings simultaneously.

Several approaches have been proposed to tackle class imbalance in long-tailed settings. Class-balanced loss~\cite{cui2019class} re-weights sample contributions based on the effective number of samples per class. Asymmetric Loss~\cite{ridnik2021asymmetric}, designed for multi-label classification, applies different focusing parameters for positive and negative samples to down-weight easy negatives. Label-Distribution-Aware Margin (LDAM) loss~\cite{cao2019learning} enforces larger decision margins for minority classes, and when combined with deferred re-weighting (DRW), applies class-balanced weights only after the network has learnt initial feature representations. Beyond loss design, two-stage training that decouples representation learning from classifier optimisation has shown benefits~\cite{kang2020decoupling}, and graph convolutional networks have been explored for modelling label co-occurrence in CXR~\cite{chen2020label}. In the medical imaging domain, recent works have investigated memory-augmented architectures~\cite{duy2025memory}, counterfactual augmentation~\cite{baek2025counterfactual} and improved ranking losses~\cite{hanif2025ranking}. The CXR-LT challenge series~\cite{lin2025cxrlt} has been instrumental in benchmarking these techniques on real-world long-tailed thoracic data.

Despite these individual advancements, there is a lack of systematic evaluation that jointly examines the interplay between loss function choice, backbone architecture capacity and post-training strategies specifically for long-tailed multi-label CXR classification. Most prior works tend to evaluate these components in isolation making it hard for practitioners to determine which combinations are most effective. The CXR-LT 2026 Challenge~\cite{dong2026overview}, built on the PadChest dataset~\cite{bustos2020padchest} with over 160K CXR images across 30 disease labels, provides an ideal testbed for such a study. Our aim is to conduct a thorough empirical evaluation investigating how loss functions, architecture selection, classifier re-training (cRT), test-time augmentation (TTA) and model ensembling interact under long-tailed conditions. Our key contributions are as follows:
\begin{enumerate}
    \item Systematic comparison of loss functions (BCE, Asymmetric, LDAM-DRW) across multiple CNN architectures from ResNet to ConvNeXt, showing that LDAM-DRW consistently benefits rare class recognition
    \item Evidence that modern architectures, specially ConvNeXt, achieve substantially better performance than conventional backbones, with ConvNeXt-Large attaining the best single-model mAP of 0.5220
    \item Evaluation of post-training strategies including cRT, TTA, calibration and ensembling, analysing their effects on both ranking and instance-level metrics
    \item A candid analysis of our 5th place submission to the CXR-LT 2026 test leaderboard, including an examination of the development-to-test performance gap
\end{enumerate}

\section{Methodology}
\label{sec:method}

\subsection{Problem Formulation}

Given a chest X-ray image $\mathbf{x}$, the goal is to predict a binary label vector $\mathbf{y} \in \{0,1\}^{C}$ where $C=30$ represents the number of disease classes. This multi-label setting allows multiple findings to co-occur in a single image. The evaluation metric is macro-averaged mAP, computed as the mean of per-class Average Precision scores.

\subsection{Loss Functions for Long-Tailed Data}

We investigated several loss functions designed to handle class imbalance:

\textbf{LDAM-DRW Loss.} The Label-Distribution-Aware Margin loss~\cite{cao2019learning} enforces larger margins for minority classes:
\begin{equation}
\mathcal{L}_{\text{LDAM}} = -\log \frac{e^{z_y - \Delta_y}}{e^{z_y - \Delta_y} + \sum_{j \neq y} e^{z_j}}
\end{equation}
where $\Delta_j = C / n_j^{1/4}$ and $n_j$ is the number of samples for class $j$. While the original formulation targets single-label softmax classification, we adapt it for the multi-label setting by applying per-class LDAM margins within a binary cross-entropy framework, treating each of the 30 classes as an independent binary classification. In the deferred re-weighting (DRW) scheme, training begins with uniform weights and transitions to class-balanced weights after an initial warm-up period, allowing the model to first learn general representations before focusing on the tail classes.

\textbf{Asymmetric Loss.} Designed for multi-label classification~\cite{ridnik2021asymmetric}, this loss applies asymmetric focusing:
\begin{equation}
\mathcal{L}_{\text{ASL}} = \sum_{i} \left[ y_i L_+ + (1-y_i) L_- \right]
\end{equation}
where $L_+$ and $L_-$ use different focusing parameters $\gamma_+$ and $\gamma_-$, with probability shifting applied to negative samples.

\subsection{Network Architectures}
We evaluate several CNN architectures spanning different design philosophies and model capacities\footnote{Values in parentheses denote approximate parameter count in millions.}: ResNet-50 (25M) and ResNet-101 (44M)~\cite{he2016deep} as foundational baselines with residual connections; DenseNet-121 (8M) and DenseNet-169 (14M)~\cite{huang2017densely} for their dense connectivity and feature reuse, which has been widely adopted in CXR analysis; EfficientFormerV2-S (22M)~\cite{tan2021efficientnetv2} as a parameter-efficient alternative; and ConvNeXt-Base (89M) and ConvNeXt-Large (198M)~\cite{liu2022convnet}, which modernise the pure CNN paradigm by incorporating Transformer-inspired design principles like patchified stems, large kernels and layer normalisation. All models were initialised with ImageNet-pretrained weights and equipped with a classification head consisting of global average pooling, layer normalisation, dropout and a linear layer outputting 30 logits.

\subsection{Post-Training Strategies: cRT, TTA and Ensemble}

Following Kang et al.~\cite{kang2020decoupling}, we explored classifier re-training (cRT) as a two-stage approach. In the first stage, the full network is trained end-to-end with LDAM-DRW loss. In the second stage, we freeze the backbone and re-initialise only the classifier head, which is then trained with class-balanced sampling to better handle infrequent classes. This effectively decouples representation learning from classifier optimisation. For test-time augmentation (TTA), we average predictions over augmented versions of each test image using horizontal flipping and small rotations ($\pm 5^{\circ}$). For ensembling, we combine predictions from multiple models through weighted averaging, assigning higher weights to models with better validation performance.

\section{Experiments}
\label{sec:experiments}

\subsection{Dataset and Preprocessing}

The CXR-LT 2026 dataset is derived from PadChest~\cite{bustos2020padchest}, comprising 160,845 chest X-ray images annotated with 30 disease labels. Following the challenge guidelines, we removed 17,933 images flagged in the official removal list, resulting in approximately 142,912 usable images. A patient-level 90/10 split was performed using GroupShuffleSplit to create internal training and validation sets, ensuring no patient data leakage. The grayscale images were replicated across three channels for compatibility with ImageNet-pretrained backbones, resized to $512 \times 512$ pixels and normalised using ImageNet statistics.

\subsection{Training Details}

All models were trained using the AdamW optimiser with a learning rate of $10^{-4}$, weight decay of $10^{-5}$, and a batch size of 8 (limited by single-GPU memory). A cosine annealing learning rate schedule was used over 50 epochs. For LDAM-DRW, deferred re-weighting was activated after 60\% of the total training epochs, with class-balanced weights derived from inverse class frequencies. Mixed precision training (FP16) was employed to reduce memory consumption and accelerate training on a single NVIDIA RTX 4050 GPU with 6GB VRAM.

\section{Results and Discussion}

\begin{table*}[htbp]
\centering
\small
\caption{Development set performance comparison across architectures, loss functions, and post-training strategies. 
Best AP and F1 are bolded.}
\label{tab:results_overview}
\setlength{\tabcolsep}{6pt}
\begin{tabular}{lllc
                S[table-format=1.4]
                S[table-format=1.4]
                S[table-format=1.4]
                S[table-format=1.4]}
\toprule
\textbf{Model} & \textbf{Loss} & \textbf{Post-training} & 
& {\textbf{AP}} & {\textbf{AUC}} & {\textbf{F1}} & {\textbf{ECE}} \\
\midrule

ResNet-50 & BCE & -- & & 0.3248 & 0.8410 & 0.3222 & 0.8884 \\
ResNet-50 & Asymmetric & -- & & 0.0667 & 0.5603 & 0.0843 & 0.9526 \\
ResNet-50 & LDAM+DRW & -- & & 0.4241 & 0.8435 & 0.2676 & 0.5575 \\
ResNet-50 & LDAM+DRW & cRT & & 0.4303 & 0.8828 & 0.3233 & 0.8300 \\
ResNet-50 & LDAM+DRW & cRT + TTA & & 0.4325 & 0.8864 & 0.3102 & 0.8247 \\

\midrule
ResNet-101 & LDAM+DRW & -- & & 0.4584 & 0.8679 & 0.2564 & 0.5332 \\

\midrule
DenseNet-121 & LDAM+DRW & -- & & 0.3967 & 0.8334 & 0.2119 & 0.5422 \\
DenseNet-169 & LDAM+DRW & -- & & 0.3981 & 0.8520 & 0.1819 & 0.5316 \\

\midrule
EfficientFormerV2-S & LDAM+DRW & -- & & 0.4869 & 0.8818 & 0.3161 & 0.5215 \\
EfficientFormerV2-S & LDAM+DRW & cRT & & 0.4539 & 0.8948 & 0.2974 & 0.8250 \\

\midrule
ConvNeXt-Base & LDAM+DRW & -- & & 0.4855 & 0.8931 & 0.3081 & 0.5319 \\
ConvNeXt-Base & LDAM+DRW & cRT & & 0.5039 & 0.8902 & 0.2548 & 0.8932 \\
ConvNeXt-Base & LDAM+DRW & cRT + TTA & & \textbf{0.5217} & 0.8961 & 0.2659 & 0.8909 \\
ConvNeXt-Base & LDAM+DRW & cRT + Prob Calib. & & 0.4539 & 0.8948 & 0.2974 & 0.8250 \\

\midrule
ConvNeXt-Large & LDAM+DRW & -- & & \textbf{0.5220} & 0.8832 & \textbf{0.3765} & 0.5506 \\
ConvNeXt-Large & LDAM+DRW & Prob Calib. & & 0.5116 & 0.8939 & 0.3669 & 0.5488 \\

\midrule
ConvNeXt-Large + EfficientFormerV2-S 
& LDAM+DRW & cRT + Ensemble & & 0.4990 & 0.8951 & 0.2556 & 0.7037 \\

\bottomrule
\end{tabular}
\end{table*}

\textbf{Effect of loss function.}
As shown in Table~\ref{tab:results_overview}, the choice of loss function has a substantial impact on performance. With ResNet-50, switching from BCE to LDAM+DRW improves mAP from 0.3248 to 0.4241, a relative improvement of over 30\%. Notably, Asymmetric Loss performs very poorly in our setting (0.0667 mAP), which we believe is due to the extreme class imbalance in CXR-LT where the default focusing parameters ($\gamma_+=0$, $\gamma_-=4$) excessively suppress gradients from the already sparse positive labels, causing the model to predict near-zero probabilities for most classes. LDAM+DRW consistently proves to be the most reliable loss across all the architectures we tested.

\textbf{Effect of architecture.}
Architectural choice also has a dominant influence on performance. Conventional CNN backbones such as ResNet and DenseNet benefit from LDAM + DRW but show limited gains in F1 as model capacity increases, indicating constrained representation learning for rare classes. In contrast, modern architectures exhibit clearer advantages: EfficientFormerV2-S achieves strong AP with comparatively low calibration error, highlighting the effectiveness of parameter-efficient designs, while ConvNeXt consistently delivers the highest AP and F1 scores. ConvNeXt-Large emerges as the strongest single model, suggesting that increased capacity combined with contemporary CNN design principles is specially beneficial under long-tailed supervision.

\textbf{Effect of classifier re-training (cRT).}
Applying classifier re-training generally improves ranking-based metrics, most notably AUC, across multiple architectures. This effect is most evident for ResNet-50 and ConvNeXt-Base, where cRT enhances decision boundaries for minority classes by decoupling representation learning from classifier optimization. However, these gains do not consistently translate to higher F1 scores and are often accompanied by increased calibration error, indicating that cRT primarily improves class separability rather than probability quality.

\textbf{Effect of TTA and ensembling.}
Test-time augmentation provides modest but consistent improvements in AP and AUC, particularly for ConvNeXt-Base, suggesting that averaging predictions over simple geometric transformations stabilizes model outputs. Nevertheless, TTA does not uniformly improve F1 and tends to worsen calibration. Ensembling further enhances AUC but fails to surpass the best single-model performance in AP or F1, reflecting a trade-off between improved ranking robustness and instance-level classification accuracy.

\begin{figure*}[htbp]
  \centering
  \includegraphics[width=0.8\linewidth]{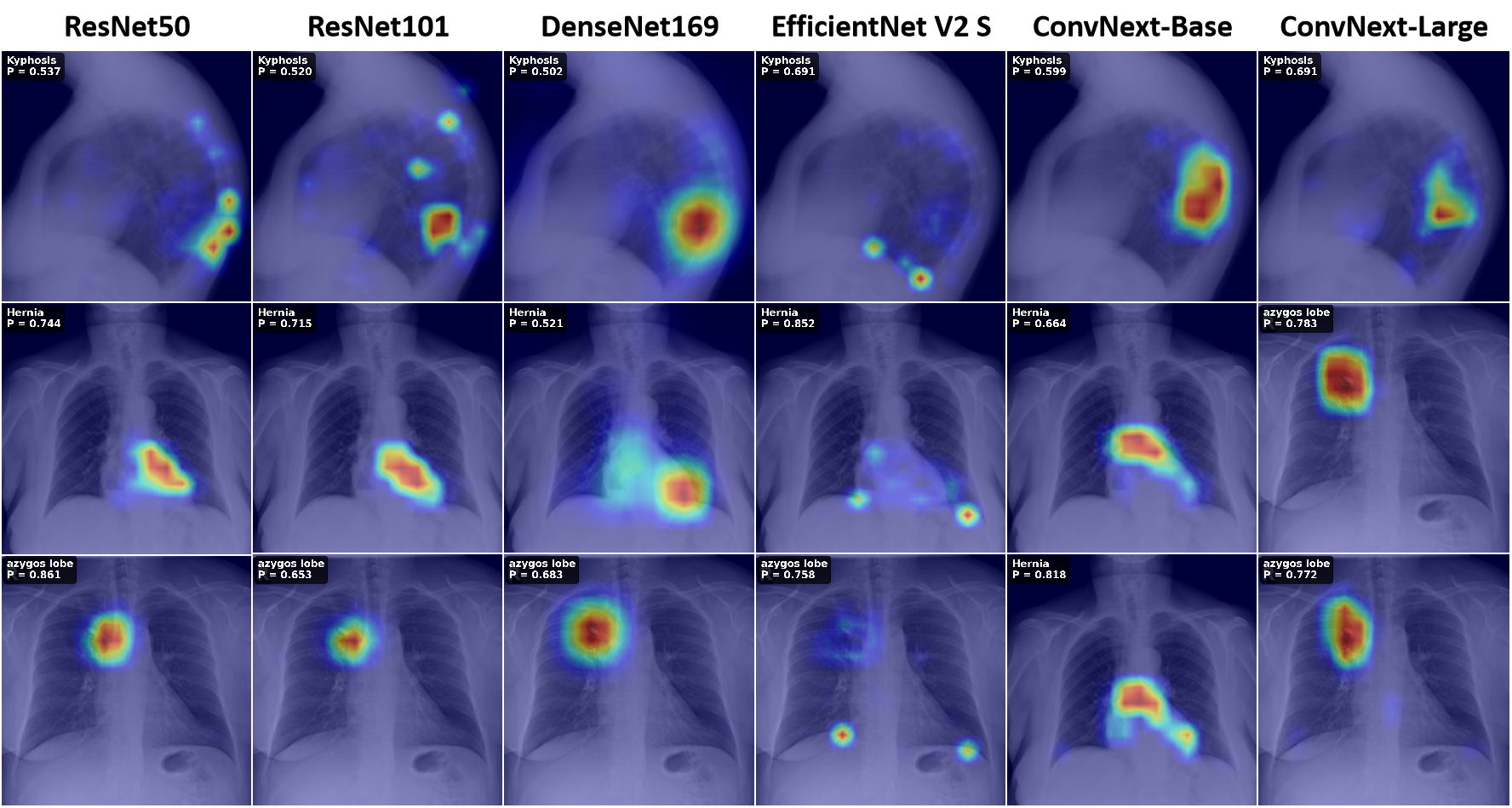} 
  \caption{Class-activation maps overlaid on CXR test images (predicted label and probability shown). The model localizes many findings correctly but probability calibration and thresholding cause instance-level misses. Rows correspond to different findings (kyphosis, hernia, azygos lobe).}
  \label{fig:qualitative_examples}
\end{figure*}

Figure~\ref{fig:qualitative_examples} presents qualitative class-activation visualizations from the test set. The model frequently places strong attention on clinically relevant regions, which aligns with its competitive AUC and mAP. However, many correctly localized predictions have probabilities near the decision threshold, resulting in instance-level misses and the substantially lower F1. These examples motivate per-class threshold tuning and stronger probability calibration (e.g., temperature scaling or isotonic regression) as immediate next steps to translate good ranking performance into improved instance-level detection.

On the official CXR-LT 2026 test leaderboard~\cite{dong2026overview}, our submission ranked 5th with a macro-averaged mAP of 0.3950, AUC = 0.8591 and F1 = 0.0945 (see Table~\ref{tab:test_leaderboard}). The top-performing team, CVMAIL x MIHL~\cite{pham2026handling}, achieved mAP of 0.5854 and AUC of 0.9259. Whilst our system showed reasonable ranking performance, the F1 score was substantially lower. The considerable drop from development mAP (0.52) to test mAP (0.395) points towards a combination of overfitting to our internal validation split, sub-optimal probability calibration and threshold selection, and ensembling choices that were more tuned towards ranking metrics rather than instance-level predictions.

\begin{table*}[htbp]
\centering
\small
\caption{Official CXR-LT 2026 Test Leaderboard (Task 1: In-distribution Multi-label Classification). 
Primary ranking metric: macro-averaged mAP.}
\label{tab:test_leaderboard}
\begin{tabular}{cllccc}
\toprule
Rank & Team & Affiliation & mAP $\uparrow$ & AUC $\uparrow$ & F1 $\uparrow$ \\
\midrule
1 & CVMAIL x MIHL & Vietnam National University, Vietnam & \textbf{0.5854} & \textbf{0.9259} & 0.3518 \\
2 & Cool Peace & KAIST Graduate School of AI, Korea & 0.4827 & 0.9186 & 0.3162 \\
3 & VIU & Vietnam National University, Vietnam & 0.4599 & 0.8827 & \textbf{0.4504} \\
4 & Bibimbap-Bueno & Case Western Reserve University, USA & 0.4297 & 0.8753 & 0.2482 \\
5 & \textbf{Nikhil Rao Sulake} & RGUKT Nuzvid, India & 0.3950 & 0.8591 & 0.0945 \\
6 & UGIVIA team & Universitat de les Illes Balears, Spain & 0.2362 & 0.7756 & 0.2353 \\
7 & 2025110451 & South-Central MINZU University, China & 0.0614 & 0.4925 & 0.0000 \\
8 & mshamani & St. Bonaventure University, USA & 0.0580 & 0.5020 & 0.0264 \\
9 & laghaei & IAU & 0.0580 & 0.5012 & 0.0302 \\
10 & uccaeid & University College Cork, Ireland & 0.0576 & 0.5001 & 0.0337 \\
\bottomrule
\end{tabular}
\end{table*}

\section{Conclusion}
\label{sec:conclusion}

We presented a systematic empirical evaluation of loss functions, architectures and post-training strategies for long-tailed multi-label chest X-ray classification on the CXR-LT 2026 benchmark. Our findings establish that LDAM-DRW loss combined with modern CNN architectures, particularly ConvNeXt, forms a strong baseline for this task, achieving 0.5220 mAP on the development set, while also revealing important trade-offs between ranking metrics and instance-level classification accuracy.

These findings bear practical significance for the deployment of automated CXR screening systems in clinical settings, where reliably detecting rare pathologies is just as critical as identifying common ones. The consistent advantage of LDAM-DRW across all architectures we tested suggests that margin-based losses with deferred re-weighting should be the default choice for clinical long-tailed tasks. At the same time, our results also show that good ranking performance alone is not sufficient, the gap between development and test mAP (0.52 vs.\ 0.395) and the very low test F1 (0.0945) highlight the pressing need for better generalisation and calibration strategies.

Going forward, per-class threshold optimisation and probability calibration (temperature scaling, isotonic regression) appears to be the most promising directions for improving instance-level predictions. Techniques like Sharpness Aware Minimisation~\cite{foret2020sharpness} and Weight Balancing~\cite{alshammari2022long} could help improve robustness to distribution shift, and graph-based methods for modelling label co-occurrence~\cite{chen2020label} may provide complementary signals to our current pipeline.

\section{Acknowledgments}
\label{sec:acknowledgments}

We thank the CXR-LT 2026 Challenge organizers for providing the benchmark dataset and evaluation platform. No external funding was received for this study.


\begingroup
\footnotesize
\setlength{\itemsep}{0pt}
\setlength{\parskip}{0pt}
\bibliographystyle{IEEEbib}
\bibliography{refs}

@article{lin2025cxrlt,
  title={{CXR-LT} 2024: A {MICCAI} challenge on long-tailed, multi-label, and zero-shot disease classification from chest X-ray},
  author={Lin, Mingquan and Holste, Gregory and Wang, Song and Zhou, Yiliang and Wei, Yuxin and Banerjee, Imon and Chen, Pingjun and Dai, Tao and Du, Yujia and Dvornek, Nicha C and others},
  journal={Medical Image Analysis},
  volume={106},
  pages={103739},
  year={2025},
  publisher={Elsevier}
}

@article{bustos2020padchest,
  title={Padchest: A large chest x-ray image dataset with multi-label annotated reports},
  author={Bustos, Aurelia and Pertusa, Antonio and Salinas, Jose-Maria and de la Iglesia-Vaya, Maria},
  journal={Medical Image Analysis},
  volume={66},
  pages={101797},
  year={2020},
  publisher={Elsevier}
}

@inproceedings{cao2019learning,
  title={Learning imbalanced datasets with label-distribution-aware margin loss},
  author={Cao, Kaidi and Wei, Colin and Gaidon, Adrien and Arechiga, Nikos and Ma, Tengyu},
  booktitle={Advances in Neural Information Processing Systems},
  volume={32},
  year={2019}
}

@inproceedings{kang2020decoupling,
  title={Decoupling representation and classifier for long-tailed recognition},
  author={Kang, Bingyi and Xie, Saining and Rohrbach, Marcus and Yan, Zhicheng and Gordo, Albert and Feng, Jiashi and Kalantidis, Yannis},
  booktitle={International Conference on Learning Representations},
  year={2020}
}

@inproceedings{cui2019class,
  title={Class-balanced loss based on effective number of samples},
  author={Cui, Yin and Jia, Menglin and Lin, Tsung-Yi and Song, Yang and Belongie, Serge},
  booktitle={IEEE/CVF Conference on Computer Vision and Pattern Recognition},
  pages={9268--9277},
  year={2019}
}

@inproceedings{ridnik2021asymmetric,
  title={Asymmetric loss for multi-label classification},
  author={Ridnik, Tal and Ben-Baruch, Emanuel and Zamir, Nir and Noy, Asaf and Friedman, Itamar and Protter, Matan and Zelnik-Manor, Lihi},
  booktitle={IEEE/CVF International Conference on Computer Vision},
  pages={82--91},
  year={2021}
}

@inproceedings{liu2022convnet,
  title={A {ConvNet} for the 2020s},
  author={Liu, Zhuang and Mao, Hanzi and Wu, Chao-Yuan and Feichtenhofer, Christoph and Darrell, Trevor and Xie, Saining},
  booktitle={IEEE/CVF Conference on Computer Vision and Pattern Recognition},
  pages={11976--11986},
  year={2022}
}

@inproceedings{tan2021efficientnetv2,
  title={{EfficientNetV2}: Smaller models and faster training},
  author={Tan, Mingxing and Le, Quoc},
  booktitle={International Conference on Machine Learning},
  pages={10096--10106},
  year={2021},
  organization={PMLR}
}

@inproceedings{he2016deep,
  title={Deep residual learning for image recognition},
  author={He, Kaiming and Zhang, Xiangyu and Ren, Shaoqing and Sun, Jian},
  booktitle={IEEE Conference on Computer Vision and Pattern Recognition},
  pages={770--778},
  year={2016}
}

@inproceedings{huang2017densely,
  title={Densely connected convolutional networks},
  author={Huang, Gao and Liu, Zhuang and Van Der Maaten, Laurens and Weinberger, Kilian Q},
  booktitle={IEEE Conference on Computer Vision and Pattern Recognition},
  pages={4700--4708},
  year={2017}
}

@article{chen2020label,
  title={Label co-occurrence learning with graph convolutional networks for multi-label chest x-ray image classification},
  author={Chen, Bingzhi and Li, Jinxing and Lu, Guangming and Yu, Hongbing and Zhang, David},
  journal={IEEE Journal of Biomedical and Health Informatics},
  volume={24},
  number={8},
  pages={2292--2302},
  year={2020},
  publisher={IEEE}
}

@article{hanif2025ranking,
  title={Enhancing multi-label chest X-ray classification using an improved ranking loss},
  author={Hanif, Muhammad Shafique and Bilal, Muhammad and Alsaggaf, Aisha H and Al-Saggaf, Umar M},
  journal={Bioengineering},
  volume={12},
  number={6},
  pages={593},
  year={2025},
  publisher={MDPI}
}

@inproceedings{baek2025counterfactual,
  title={Counterfactual augmentation for long-tailed multi-label chest X-ray classification},
  author={Baek, Seunghyun and Shin, Jitae},
  booktitle={International Workshop on Efficient Medical Artificial Intelligence},
  pages={357--366},
  year={2025},
  publisher={Springer}
}

@inproceedings{duy2025memory,
  title={Mitigating class imbalance in chest X-ray classification with memory-augmented models},
  author={Duy, Hoang Khoi and Duy, Nguyen Huu and Ngu, Hoang Cong Viet},
  booktitle={24th International Conference on Computational Science},
  year={2025},
  publisher={IEEE}
}

@inproceedings{alshammari2022long,
  title={Long-tailed recognition via weight balancing},
  author={Alshammari, Shaden and Wang, Yu-Xiong and Ramanan, Deva and Kong, Shu},
  booktitle={Proceedings of the IEEE/CVF conference on computer vision and pattern recognition},
  pages={6897--6907},
  year={2022}
}

@article{foret2020sharpness,
  title={Sharpness-aware minimization for efficiently improving generalization},
  author={Foret, Pierre and Kleiner, Ariel and Mobahi, Hossein and Neyshabur, Behnam},
  journal={arXiv preprint arXiv:2010.01412},
  year={2020}
}

@article{pham2026handling,
  title={Handling Supervision Scarcity in Chest X-ray Classification: Long-Tailed and Zero-Shot Learning},
  author={Pham, Ha-Hieu and Nguyen, Hai-Dang and Nguyen, Thanh-Huy and Xu, Min and Bagci, Ulas and Le, Trung-Nghia and Pham, Huy-Hieu},
  journal={arXiv preprint arXiv:2602.13430},
  year={2026}
}

@article{dong2026overview,
  title={Overview of the CXR-LT 2026 Challenge: Multi-Center Long-Tailed and Zero Shot Chest X-ray Classification},
  author={Dong, Hexin and Lin, Yi and Zhou, Pengyu and Zhao, Fengnian and Legasto, Alan Clint and Lin, Mingquan and Chen, Hao and Yang, Yuzhe and Shih, George and Peng, Yifan},
  journal={arXiv preprint arXiv:2602.22092},
  year={2026}
}
\endgroup

\end{document}